\documentclass[10pt]{article}
\usepackage[utf8]{inputenc}
\usepackage[T1]{fontenc}
\usepackage{amsmath ,amssymb ,amsfonts,mathrsfs,wasysym,amsthm, textcomp, mathtools,graphicx,stmaryrd}
\DeclareMathAlphabet{\mathpzc}{OT1}{pzc}{m}{it}
\graphicspath{ {/home/elqrq/Desktop/Tesi/Img/} }

\newcommand{\A}{\mathcal{A}}

\newcommand{\Rea}{\mathbb{R}}
\newcommand{\Comp}{\mathbb{C}}

\newcommand{\Nat}{\mathbb{N}}

\newcommand{\Ho}{\mathcal{H}_\omega}
\newcommand{\Hi}{\mathcal{H}}

\newcommand{\st}{\mspace{5mu} | \mspace{5mu} }

\newcommand{\field}{\mathbb{K}}

\newcommand{\boundo}{\mathcal{B}(\mathcal{H})}

\newcommand{\boundoo}{\mathcal{B}(\mathcal{H}_{\omega})}

\newcommand{\borel}{\mathscr{B}(\mathbb{R})}

\newcommand{\borelx}{\mathscr{B}(\mathsf{X})}

\newcommand{\unit}{\mathsf{I}}

\newcommand{\Tr}[1]{\mbox{Tr}\left[#1\right]}

\newcommand{\X}{\mathsf{X}}

\begin{document}

	\newtheorem{definition}{Definition}
	\newtheorem{proposition}{Proposition}
	\newtheorem{theorem}{Theorem}
	\newtheorem{lemma}{Lemma}
	\newtheorem{corollary}{Corollary}
	\newtheorem{assumptions}{Assumptions}
	\newtheorem{assumption}{Assumption}

	\author{Curcuraci Luca \\ Department of Physics, University of Trieste, \\ Strada Costiera 11 34151, Trieste, Italy \\
		\small {Curcuraci.article@protonmail.com} \\
			\small{luca.curcuraci@phd.units.it} }
	\title{\huge{Why do we need Hilbert spaces?} }
	\date{March 2016}
	\maketitle

	\begin{abstract}
		These are the notes written for the talk given at the workshop \textquotedblleft \emph{Rethinking foundations of physics 2016}". In section \ref{sec2}, a derivation of the the quantum formalism starting from propositional calculus (quantum logic) is reviewed, pointing out which are the basic requirements that lead to the use of Hilbert spaces. In section \ref{sec3}, a similar analysis is done following for the reconstruction of quantum theory using an operational approach. In both cases, non-commutativity plays a crucial role. Finally, in section \ref{sec4} a toy model which try to motivate non-commutativity is proposed. Despite this last section is interesting to read, the analysis performed there is not complete. This toy model will be re-formulated in a rigorous way (and extended) in future works.
	\end{abstract}
	\newpage
	\tableofcontents
	\newpage
	
	\section{Introduction}

	Despite all its success, quantum mechanics after more that one hundred years, is still under debate. All the problems due to its interpretation are originated from the choice to formulate the theory using the Hilbert space formalism. Here we will discuss how this choice is unavoidable and suggests an interesting motivation for this unavoidability.\newline
	
	In general, two are the main basic observations that can be done, experimentally, when one deal with non-relativistic quantum systems
	\begin{enumerate}
		\item[\textbf{O1}] the outcomes of an experiment about a quantum system is probabilistic;	
		\item[\textbf{O2}] there are physical quantities that can be measured \emph{simultaneously} and other that do not.
	\end{enumerate}
	The first observation is a common feature of all the physical system, once one take seriously into account the fact that, even the best experimental physicist in the word, can  perform measurement with a finite resolution. The second observation is the distintive feature of a quantum system and, as we shall see, it is the origin of all the differences between classical and quantum system. Starting from these two observations, one can derive (as logical consequence and with few further assumptions) almost all the postulates of quantum mechanics. In what follow we will present two approaches for such derivation, pointing out where we need to do an assumption in order to continue the reconstruction. The first approach is based on (quantum) logical arguments, while the second make use of operational considerations. Finally we will present a toy model to motivate the second observation (deriving it from the first, in some sense) for the description of a point-like particle, moving over a random space.

	\section{Propositions about a quantum system: QM from QL}\label{sec2}

	In this section, the quantum logic derivation of the Hilbert space structure of quantum mechanics will be briefly reviewed. The main idea behind this approach is to find the theoretical foundations of the postulates for a quantum theory starting from the proposition that one can formulate about a quantum system. This is the so called quantum logic (QL) approach to the foundations of quantum mechanics (QM). For a more detailed treatment, we refer to \cite{gBjvN},\cite{eBgC}, \cite{mpS} for the quantum logic and \cite{vM} for the mathematical formulation of the quantum mechanics' postulates. \newline
	
	\subsection{Propositions for quantum systems}
	
	In everyday life, it is a common fact to formulate \emph{propositions} to describe something and this, of course, holds also in science. Propositions are the basic outcomes of any experiment, and so it is reasonable to expect that some basic feature of the physical system under study, can be deduced from the propositions we may formulate from the experiments. If  a quantity $A$ can be measured (assign an objective numerical value), it is common to formulate proposition like
	\begin{center}
		\emph{\textgravedbl$A$ takes the value $v_A$\textacutedbl}
	\end{center}
	or better, taking into account the finite resolution of any measurement device (hence O1)
	\begin{center}
		\emph{\textgravedbl$A$ takes value in $[a,b]$\textacutedbl}
	\end{center}
	These proposition can be considered as the simplest possible proposition. Introducing the very natural logical connectivities AND/OR
	and considering a second measurable quantity $B$, one may also formulate composite propositions. Two basic examples are
	\begin{center}
		\emph{\textgravedbl$A$ takes value $v_A$OR $B$ takes value $v_B$\textacutedbl}\\
		\emph{\textgravedbl$A$ takes value $v_A$ AND $B$ takes value $v_B$\textacutedbl}
	\end{center}
	In everyday life, both the propositions make sense. Nevertheless, for a quantum system, the second proposition cannot be formulated in general: if $A$ and $B$ cannot be measured simultaneously, the measurement to formulate this proposition cannot be performed in general because \textbf{O2}. We can see that the effect of \textbf{O2} is to reduce the number of propositions that we may formulate using AND. Let us also observe the following fact: the proposition
	\begin{center}
		\emph{\textgravedbl A take value $v_A$ IMPLIES THAT $B$ take value $v_B$\textacutedbl}
	\end{center}
	make sense only for quantities that can be measured at the same time, which is again a consequence of \textbf{O2}. In order to explore better the consequences of this, we will adopt the following conventions: a proposition like \emph{\textgravedbl $A$ takes value $v_A$\textacutedbl} or \emph{\textgravedbl $A$ takes value $[a,b]$\textacutedbl} will be labeled simply by $a$, the logical connectors AND by $\wedge$, $OR$ by $\vee$, the implication by $\Rightarrow$ and finally the logical negation by $\neg$. We say that two propositions are equal $a = b$ when $a \Rightarrow b$ and $b \Rightarrow a$. With this notation, the observation \textbf{O2} restrict the number of propositions about a physical system having form $a \wedge b$ which make sense. Before to go on, suppose $X$ may take only two values $v_X$ and $u_Y$. Then consider the following proposition
	\begin{center}
		\emph{\textgravedbl $Y$ takes value in $v_Y$ AND, $X$ take value $v_X$ OR $X$ take value $u_X$\textacutedbl}
	\end{center}
	Apparently it seems equivalent to
	\begin{center}
		\emph{\textgravedbl $Y$ takes value in $v_Y$ AND $X$ takes value $v_X$, OR $Y$ takes value in $v_Y$ AND $X$ takes value $u_X$\textacutedbl}
	\end{center}
	Using the conventions introduced above $a \wedge (b \vee c) = (a \wedge b) \vee (a \wedge c)$. It is not difficult to understand that if $X$ and $Y$ cannot be measured at the same time, the second proposition do not make sense: this means that for the possible propositions that we can formulate about a quantum system, in general 
	\begin{equation}\label{dist}
		a \wedge (b \vee c) \neq (a \wedge b) \vee (a \wedge c)
	\end{equation}
	where $\neq$ simply means that it is not true that are equivalent. Consider now a different situation. Let $X,Y,Z$ be three measurable quantities. If
	\begin{center}
		\emph{\textgravedbl $X$ takes value $v_X$ IMPLIES THAT $Y$ takes value $v_Y$\textacutedbl}
	\end{center}
	is true, namely the value assumed by $X$ determine the value of $Y$, then
	\begin{center}
		\emph{\textgravedbl$X$ takes value $v_X$, OR $Y$ takes value $u_Y$ AND $Z$ takes value $w_Z$ IMPLIES THAT $Y$ takes value $u_Y$, AND $X$ takes value $v_X$ OR
			$Z$ takes value $w_Z$\textacutedbl}
	\end{center}
	which in symbols can be written as: if $a \Rightarrow b$ then $a \vee (b \wedge c) \Rightarrow b \wedge (a \vee c)$. Again we can see that, if $a$ and $c$ cannot be measured at the same time (and so also $b$ cannot be measured at the same time of $c$), the first part of this proposition in general doesn't make sense. This means that for the possible propositions we can formulate about a quantum system, in general
	\begin{equation}\label{modu}
		\mbox{if } a \Rightarrow b \mbox{ then } a \vee (b \wedge c) \nRightarrow b \wedge (a \vee c)
	\end{equation}
	This is again a consequence of \textbf{O2}. Finally we observe that in general when 
	\begin{center}
		\emph{\textgravedbl $X$ takes value $v_X$ IMPLIES THAT $Y$ takes value $v_Y$\textacutedbl}
	\end{center}
	is true, for $u_Y$ arbitrary, the proposition
	\begin{center}
		\emph{\textgravedbl$X$ takes value $v_X$, OR $X$ DOES NOT take value $v_X$ AND $Y$ takes value $u_Y$ IS EQUIVALENT TO $Y$ takes value $u_Y$\textacutedbl}
	\end{center}
	always make sense for a quantum system, since the two physical quantities can always be measured at the same time by assumption. In symbols we can write that, for the set of propositions about a quantum system
	\begin{equation}\label{othomodu}
		\mbox{if } a \Rightarrow b \mbox{ then } a \vee (\neg a \wedge b) = b 
	\end{equation}
	holds. From this discussion we can understand the the logical connectors OR, AND, NOT, IS EQUIVALENT TO and IMPLIES THAT cannot be used in a straightforward manner for a quantum system: thus the usual logic is not suitable in this case. As we will see, (\ref{dist}),(\ref{modu}) and (\ref{othomodu}) will help us to select the right structure to describe mathematically the set of all the propositions we may formulate about a quantum system, namely to implement \textbf{O2}.
	
	\subsection{From propositions to lattice}
	
	Let us now try to formalise mathematically the discussion done before. In order to do that, we need to state some technical definitions.
	\begin{definition}
		Let $X$ be a set. A relation $\preccurlyeq$ on $X$ is said \emph{partial order} if it is reflexive ($x\preccurlyeq x$, $\forall x \in X$), transitive ($x \preccurlyeq y$ and $y \preccurlyeq z$ implies $x \preccurlyeq z$, $\forall x,y,z \in X$) and skew-symmetric ($x \preccurlyeq y \preccurlyeq x$ implies $x = y$, $\forall x,y \in X$). The couple
		$(X,\preccurlyeq)$ is said \emph{poset}. 
	\end{definition}
	Using the ordering relation of the poset, one may define the following
	\begin{definition}
		Let $(X,\preccurlyeq)$ be a poset and consider a subset $Y \subset X$. The \emph{lower bound of $Y$} (\emph{upper bound of $Y$}) is an element $a \in X$ such that $a \preccurlyeq x$ ($x \preccurlyeq a$) for any $x \in X$. The \emph{greatest lower bound, GLB} (\emph{least upper bound, LUB}) of $Y$ is a lower bound (upper bound) of $Y$ $b$ such that $b \preccurlyeq a$ ($a \preccurlyeq b$) for every lower bound (upper bound) $a$ of $Y$.
	\end{definition}
	It is not difficult to see that if the GLB (LUB) exists it is unique. Now we are ready to introduce the central mathematical concept of this paragraph.
	\begin{definition}
		Given a poset $(X,\preccurlyeq)$, it is a \emph{lattice} if for any $x,y \in X$, the GLB and LUB always exist (denoted $x \wedge y$ and $x \vee y$, respectively). 
	\end{definition}
	Not all the poset are lattice. The symbols $\wedge$ and $\vee$ used in the definition above, can be defined as the following maps
	\begin{enumerate}
		\item[a)] $\wedge: X \times X \rightarrow X$, such that for any $x,y,z \in X$, then $x \wedge y \preccurlyeq x$, $x \wedge y \preccurlyeq y$ and, if $z \preccurlyeq x$ and $z \preccurlyeq y$ then $z \preccurlyeq x \wedge y$.
		\item[a)] $\vee: X \times X \rightarrow X$, such that for any $x,y,z \in X$, then $x \preccurlyeq x \vee y$, $y \preccurlyeq x \vee y$ and, if $x \preccurlyeq z$ and $y \preccurlyeq z$ then $x \wedge y \preccurlyeq z$.
	\end{enumerate}
	and it is not difficult to see that the writing $x \wedge y = x$, $x \vee y = y$ and $x \preccurlyeq y$ are equivalent. Lattices are classified according to the following
	\begin{definition}
		A lattice $(X,\preccurlyeq)$ is said
		\begin{enumerate}
			\item[a)] \emph{distributive} if $x \wedge (y \vee z) = (x \wedge y) \vee (x \wedge z)$, $\forall x,y,z \in X$;
			\item[b)] \emph{modular} if $x \preccurlyeq y$ implies $x \vee (y \wedge z) = y \wedge (x \vee z)$, $\forall x,y,z \in X$;
			\item[c)] \emph{bounded} if there exist two elements $\mathbf{0} \in X$ and $\mathbf{1} \in X$ such that $\mathbf{0} \preccurlyeq x \preccurlyeq \mathbf{1}$, $\forall
			x \in X$;
			\item[d)] \emph{orthocomplemented} if it is bounded and equipped with an operation $x \mapsto \neg x$ (called \emph{orthocomplementation}) such that
			\begin{enumerate}
				\item[i)] $x \vee \neg x = \mathbf{1}$, $\forall x \in X$;
				\item[ii)] $x \wedge \neg x = \mathbf{0}$, $\forall x \in X$;
				\item[iii)] $\neg (\neg x) = x$, $\forall x \in X$;
				\item[iv)] $x \preccurlyeq y$ implies $\neg y \preccurlyeq \neg x$, $\forall x,y \in X$;
			\end{enumerate}
			\item[e)] \emph{orthomodular} if orthocomplemented and $x \preccurlyeq y$ implies that $x \vee (\neg x \wedge y) = y$, $\forall x,y \in X$;
		\end{enumerate}
	\end{definition}
	One can prove that: distributivity implies modularity  which implies orthomodularity, but the converse is not true. The last notions we need to reach the goal of this paragraph, are about the elements of a lattice
	\begin{definition}
		Let $(X,\preccurlyeq)$ be a bounded lattice then
		\begin{enumerate}
			\item[a)] an element $x \in X$ \emph{covers} $y \in X$ if $y \prec x$ (namely, $y \preccurlyeq x$ but $x \neq y$) and doesn't exist $z \in X$, such that $y \prec z \prec x$;
			\item[b)] an element $x \in X$ is said \emph{atom} if it covers $\mathbf{0}$;
			\item[c)] two elements $x,y \in X$ are said \emph{orthogonal}, written $x \perp y$, if $x \preccurlyeq \neg y$
		\end{enumerate}
		A bounded lattice $(X,\preccurlyeq)$ is said \emph{atomic} if for any $y \in X/\{\mathbf{0}\}$ there exist an atom $x \in X$ such that $x \preccurlyeq y$. A bounded lattice is said \emph{atomistic} if any element of the lattice can be seen as the join of atoms. An atomic lattice $(X,\preccurlyeq)$ is said \emph{with the covering property} if for any $x \in X$ and every atom $a \in X$ such that $a \wedge x = \mathbf{0}$, the element $a \vee x$ covers $x$.
	\end{definition}
	For an orthomodular lattice, one can prove that if it is atomic, it is also atomistic. The discussion done in the previous paragraph seems to suggest the following: if the set of all the propositions about a quantum system with the operations $\Rightarrow, \wedge$ and $\vee$ is a lattice, then it must be an orthomodular lattice because of \textbf{O2}. Nevertheless, in order to say this we need to find a way to define the partial ordering, namely the $\Rightarrow$ that in the previous paragraph played the role of logical implication. To define this ordering relation, the observation \textbf{O1}, suggests that the following mathematical definition is physically reasonable
	\begin{definition}
		Let $(X,\preccurlyeq)$ be an orthomodular lattice, a \emph{probability-like measure} on $X$ is a function $p: X \rightarrow [0,1]$ such that
		\begin{enumerate}
			\item[a)] $p(\mathbf{1}) = 1$ and $p(\mathbf{0}) = 0$;
			\item[b)] for every sequence $\{x_i\}_{i \in I}$ of orthogonal elements of $X$, $p\left( \bigvee_i x_i\right) = \sum_i p(x_i)$
		\end{enumerate}
	\end{definition}
	This probability-like measure induces an ordering relation on $X$, in particular $x \preccurlyeq y$ if and only if $p(x) \leqslant p(y)$ for every possible $p$: but notice that the ordering relation exists independently to the existence of $p$. In any case, the observation \textbf{O1} tells us that when we study a quantum system (and in general any physical system) this notion is at disposal: the measure $p$ can be interpreted as a \textgravedbl degree of belief\textacutedbl (or \textgravedbl truth value\textacutedbl) of a certain proposition, namely, if $a$ is a proposition, $p(a)$ tell us how much we are sure that $a$ happens in real word. But one must be careful about \textbf{O2}: the degree of belief of a proposition can be tested and compared with the one of another proposition, only if these propositions are associated to observables that are measurable at the same time. Keeping this fact in mind, we can say that, if we are agree on \textbf{O1}, we have an ordering relation at disposal over the set of all the physical propositions. This partial order, allows us to define the met and join between all the propositions (actually this is an assumption, despite it is reasonable), thus we can conclude that it is a lattice. In what follow, the lattice of the physical proposition about a quantum system $Q$ will be denoted by $\mathcal{L}_Q(\Rightarrow)$, where $\Rightarrow$ denotes the partial ordering relation described before. The observation \textbf{O2} suggests that is an orthomodular lattice because we expect (\ref{othomodu}) to hold. Nevertheless we should also check if it is bounded and define an orthocomplementation on it. 
	We need to define $\mathbf{1}$ and $\mathbf{0}$. The first can be though as the proposition
	\begin{center}
		\emph{\textgravedbl The measurement of some quantity is a real number \textacutedbl}
	\end{center}
	which is clearly always true for a physical system. $\mathbf{0}$ can be thought as the proposition
	\begin{center}
		\emph{\textgravedbl We are not measuring anything \textacutedbl}
	\end{center}
	which is always false if we assume that we formulate propositions only after that at least one experiment was performed. It is not difficult to see that any other proposition in between these two, or more formally, $\mathbf{0} \Rightarrow a \Rightarrow \mathbf{1}$. Thus $\mathbf{0}$ and $\mathbf{1}$ belongs to the lattice of all the propositions about the physical system $\mathcal{L}_Q(\Rightarrow)$ and this lattice is bounded. Once we have this, the orthocomplementation of a proposition $a \in \mathcal{L}_Q(\Rightarrow)$ is the unique proposition $\neg a$ with truth value $p(\neg a) = 1-p(a)$, and it is not difficult to understand that it is the negation (in common language sense) of the initial proposition. Hence it is reasonable to think $\mathcal{L}_Q(\Rightarrow)$ as an \emph{orthomodular lattice}.
	
	Now, we will try to motivate other two properties that $\mathcal{L}_Q(\Rightarrow)$ should have: atomiticy and the covering property. When we deal with a physical system we have at disposal a set of elementary propositions, like
	\begin{center}
		\emph{\textgravedbl The physical quantity $A$ takes exactly the value $v_A \in \Rea$\textacutedbl}
	\end{center}
	and from them we may construct more complex propositions (like \textgravedbl$A$ takes value in $[a,b]$\textacutedbl). In principle for a physical system, we have at disposal an infinte number of this kind of propositions. Moreover, if $A$ is an elementary physical quantity, in the sense that cannot be expressed in terms of other physical quantities, then these propositions may cover only $\mathbf{0}$. This means that propositions of this kind correspond to the atoms of $\mathcal{L}_Q(\Rightarrow)$. It is also physically reasonable to say that, any proposition about a physical system or is an atom or it cover an atom (excluding the trivial case of the $\mathbf{0}$ proposition). Hence this suggest that $\mathcal{L}_Q(\Rightarrow)$ is an \emph{atomistic lattice}. The covering property is more subtle but still reasonable. Suppose we have two elementary propositions $a$ and $b$ (hence two atoms) and consider also a third proposition $c$ (which may not be in general an atom). Now, suppose that we know $a \Rightarrow b\vee c$. This means that $p(a) \leqslant p(b \vee c)$, and so that at the same time $a,b$ and $c$ should hold. Because of this simultaneous truth of these propositions, and because $a$ cannot implies $b$ and viceversa (they are atoms), it is also reasonable to assume that $p(b) \leqslant p(a \vee c)$ (which means $b \Rightarrow a\vee c$) simply because, if it is not so, for a physical system the previous requirement ($a \Rightarrow b\vee c$) doesn't make sense anymore. Thus more rigorously we can write, if $a,b$ are atoms, then $a \Rightarrow b\vee c$ implies $b \Rightarrow a\vee c$ for any $c$. This is another possible characterisation of the covering property for the case of orthomodular lattice. Hence $\mathcal{L}_Q(\Rightarrow)$ can also be considered as a \emph{lattice with the covering property}.
	
	Finally we conclude this paragraph with the last lattice-theoretical concept which, by the way, can always be assumed: \emph{irreducibility}.
	\begin{definition}
		Let $(X,\preccurlyeq)$ be an orthocomplemented lattice. Consider two elements $a,b \in X$, we say that $a$ \emph{commute with} $b$ if 
		\begin{equation*}
			a = (a \wedge b) \vee (a \wedge \neg b)
		\end{equation*}
		The set of all the elements of the lattice commuting with any other element of the lattice is called \emph{center}. A lattice is said \emph{irreducible} if its center is just $\{\mathbf{0},\mathbf{1}\}$.
	\end{definition}
	We can easily see that two propositions commute if and only if they are testable at the same time (hence they are associated to two simultaneously measurable quantities or to the same quantity). Thus the commutativity can be interpreted as simultaneous testability. This means that \textbf{O2} implies the loss of commutativity between propositions, in lattice-theoretical terms, and so it determines the impossibility to use the usual interpretation of the logical connectors, as discussed in the beginning. It can be proved that any reducible (i.e. not irreducible) lattice can be seen as the direct sum (in set-theoretical sense) of lattices that are irreducible. Hence, even if the set of propositions about a quantum system is not irreducible, we may always recast the problem in lattices that are irreducible. For this reason we will always consider irreducible lattices. 
	\newline
	Thus we may conclude the following fact: \emph{the lattice of propositions we can formulate on a quantum system $Q$, ${\mathcal{L}_Q(\Rightarrow)}$, is an orthomodular, irreducible, atomistic lattice with the covering property whose ordering relation is represented by the truth value of a proposition}.
	
	\subparagraph{Remark.}
	The arguments presented here doesn't prove rigorously that a quantum system, and the set of propositions about it, are described by the lattice $\mathcal{L}_Q(\Rightarrow)$ with the properties mentioned above. The aim of this paragraph is to convince the reader that it is physically well motivated to assume this structure as starting point, and that the motivations lie at the heart of all the experimental observations about a quantum system.
	
	\subsection{From lattice to Hilbert spaces}
	
	We have seen that an orthomodular, irreducible, atomistic lattice with the covering property can be used to model the set of proposition we can formulate about a quantum system. In this paragraph we will see how it is possible to map this rather abstract mathematical structure to the usual Hilbert spaces in which quantum mechanics is typically formulated. Two are the main results that we need, but before to state them, we need to introduce some technical definitions.
	\begin{definition}
		Let $\field$ be a division ring and consider a vector space $\Hi$ on it. Then
		\begin{enumerate}
			\item[a)] an \emph{involution} is a map $^* : \field\rightarrow \field$ such that $(a+b)^* = a^* + b^*$, $(ab)^* = b^* a^*$ and $(a^*)^* = a$, $\forall a,b \in \field$;
			\item[b)] an \emph{hermitian form} is a map $\langle \cdot , \cdot \rangle : \Hi \times \Hi \rightarrow \field$ such that
			\begin{enumerate}
				\item[i)] $\langle x , y \rangle = (\langle y, x \rangle)^*$, $\forall x,y \in \Hi$;
				\item[ii)] $\langle x, ay + bz \rangle = a \langle x,y \rangle + b \langle x,z \rangle$, $\forall x,y,z \in \Hi$ and $\forall a,b \in \field$;
				\item[iii)] $\langle x, x\rangle = 0$ if and only if $x =0$, $\forall x \in \Hi$
			\end{enumerate} 
		\end{enumerate}
		The couple $(\Hi,\langle \cdot ,\cdot \rangle)$ is called \emph{hermitian inner space}.
	\end{definition}
	As usual, a subset where the vector space operation of $\Hi$ are preserved is called \emph{subspace}. As usual, two elements $x,y \in \Hi$ are said orthogonal if $\langle x , y \rangle = 0$. This allows us to define the orthogonal complement of a subspace $N \subset \Hi$, which is the set
	\begin{equation*}
		N^\perp := \{ y \in \Hi \quad | \quad \langle x , y \rangle = 0 , \forall x \in N\}
	\end{equation*}
	Then the following definition holds
	\begin{definition}
		Given an hermitian inner space $(\Hi, \langle \cdot , \cdot \rangle )$, if for any closed subspace $N \in \Hi$ one can write that $\Hi = N \oplus N^\perp$, then $(\Hi,\langle \cdot , \cdot \rangle)$ is said \emph{generalised Hilbert space} (or \emph{orthomodular space}).
	\end{definition}
	We observe that, the class of Hilbert spaces is a particular class of generalised Hilbert spaces, in fact in this definition the involution and the field $\field$ are arbitrary.
	Now, we are ready to state the first important theorem, due to Piron, that allows us to recover the Hilbert space formulation.
	\begin{theorem}[Piron theorem]
		Any (complete) irreducible, atomistic orthomodular lattice $(X,\preccurlyeq)$ with the covering property having at least four orthogonal atoms, is isomorphic to the set of closed subspaces of some generalised Hilbert space $(\Hi,\langle \cdot , \cdot \rangle)$.
	\end{theorem}
	$\mathcal{L}_Q(\Rightarrow)$ fulfil all the requirements of this theorem (the \emph{completeness} for a lattice was assumed when we declared that the met and join always exist.) except for the number of atoms which are orthogonal. Orthogonal atoms, means elementary propositions that are mutually exclusive. This means that, if we try to evaluate the first proposition, then we know that the second proposition is not true: this does not seem to be an unphysical requirement since it seems reasonable that the number of mutually exclusive propositions is infinite (because in general a physical quantity assume value over $\Rea$). So, accepting that we have at least 4 orthogonal atoms, the Piron theorem guarantees that our lattice can be represented using some generalised Hilbert space $\Hi$, and in particular the propositions are in one-to-one correspondence with the closed subspace of $\Hi$ (i.e. in one-to-one correspondence with projectors over these subspaces).
	
	We have not jet reached our goal to motivate the Hilbert space structure of quantum mechanics with the quantum logic approach. In order tho do so we need a second theorem, due to S\'{o}ler, which is able to select between all the generalised Hilbert spaces exactly the three class of Hilbert spaces.
	\begin{theorem}[S\'{o}ler theorem]
		If a generalised Hilbert space $(\Hi, \langle \cdot , \cdot \rangle)$ over a field $\field$, admits a sequence of elements $\{e_i\}_{i \in \Nat}$ such that
		\begin{equation*}
			\langle e_i , e_j \rangle = \delta_{i,j} \lambda
		\end{equation*}
		for some $\lambda \in \field$, then $\field$ must be the field of reals, complex or quaternionic numbers and  $(\Hi, \langle \cdot , \cdot \rangle)$ is an infinite dimensional Hilbert space over one of these fields.
	\end{theorem}
	Thus we only need to find a sequence $\{e_i\}_{i \in \Nat}$ of pairwise orthogonal element of $\Hi$. The existence of such sequence can be motivated from the physical assumption of an infinite number of orthogonal atoms associated to the same physical quantity: atoms are elementary propositions about this quantity, which are in one-to-one correspondence with the closed subspace of $\Hi$. Orthogonality between atoms translate in orthogonality between subspaces, thus we have an infinte number of orthogonal subspace on which a single physical quantity takes different values. This suggests that the dimension of $\Hi$ should be infinite for a reasonable physical theory. In this way, one can motivate heuristically the application of the Soler theorem. Nevertheless we still not select a particular Hilbert space. The Hilbert space over the field of real numbers can be excluded from considerations about the Galilean invariance, for a non-relativistic quantum theory. Hilbert spaces over the field of quaternionic numbers, are still under studies, but they can be seen as the direct sum of two complex Hilbert spaces. So it seem that the field of complex number is in some sense special, and it is actually the arena where standard quantum mechanics is formulated.\newline
	Despite we are not able to select uniquely the usual Hilbert space of quantum mechanics, the quantum logic approach gives a strong argument supporting the usual formulation of quantum mechanics in complex Hilbert space. One can see that, as a consequence of a very basic physical consideration (the observation \textbf{O2}), this structure cannot be avoided.
	
	\subsection{From Hilbert space to quantum mechanics}
	
	In this last paragraph, we briefly explain how it is possible to obtain all the remaining postulate of quantum mechanics, as logical consequences of the structure explained before plus some assumptions.
	
	Let us recap the results obtained in the previous paragraph. Assuming that the propositions about a quantum system fulfil \textbf{O1} and \textbf{O2}, plus some other reasonable hypothesis, we are lead to the following conclusion:
	\begin{center}
		\emph{To each quantum system we may associate a complex Hilbert space $(\Hi,\langle \cdot ,\cdot \rangle)$. Propositions about a quantum system are represented by closed subspaces of $\Hi$.}
	\end{center}
	Over this Hilbert space, then we may introduce a probability-like measure which tells us the truth value of each proposition. Such probability-like measure, by the \emph{Gleason theorem} (assuming $\dim \Hi > 2$), is uniquely determined by a positive trace-class operator $\hat{\rho}$. Such operator can be always normalised and it allows to say that the truth value of a proposition $A \subset \Hi$ si given by
	\begin{equation*}
		P(A) = \Tr{\hat{\rho}\hat{P}_A}
	\end{equation*}
	where $\hat{P}_A$ is the projector associated to the closed subspace $A$. This means that
	\begin{center}
		\emph{To each quantum system we may associate a positive, normalised, trace-class operator $\hat{\rho}$, which allows us to compute the probability to find a given proposition true as 
			$ \Tr{\hat{\rho}\hat{P}_A}$}
	\end{center}
	Typically, such positive, normalised, trace-class operator is called \emph{state}. At this point we need to define the notion of observable. Among all the propositions about a physical system, all the propositions regarding the same physical quantity should have these features
	\begin{enumerate}
		\item[1.] they are all simultaneously testable, since are all associated to the same physical quantity. This implies that all the propositions commute and so in this case the usual interpretation of logical connectors can be applied;
		\item[2.] call $m(A)$ is the outcome of a measurement of $A$, if $m(A)\in B$ is true and $m(A) \in B'$ is also true, then clearly $m(A) \in B \cap B'$ is true;
		\item[3.] the proposition $m(A) \in \Rea$ is always true, thus correspond to $\mathbf{1}$;
		\item[4.] if $m(A) \in B$ and $B$ can be written as $B = \cup_{i \in I} C_i$, then the proposition $m(A) \in B$ is equivalent to the proposition $\vee_{i \in I}\{m(A) \in C_i \}$, where $\vee$ is interpreted as OR, since from the point 1 we know that the usual interpretation of logical connectivities can be applied.
	\end{enumerate}
	Defining the set of propositions fulfilling all these physically reasonable features, one can prove that to each physical quantity it is possible to associate a projection-valued measure (PVM). Then by the \emph{spectral theorem} to each PVM one can associate a self-adjoint operator to each physical quantity. Moreover, one can always associate to PVM a probability measure which can be used to compute the expectations (the so called \emph{spectral measure}). More precisely, one can prove the following
	\begin{center}
		\emph{Physical quantities are represented by self-adjoint operators over $\Hi$. For a quantum system with state $\hat{\rho}$, the expectation value of a physical quantity represented by a self-adjoint operator $\hat{A}$ is given by $\Tr{\hat{\rho}\hat{{A}}}$}
	\end{center}
	At this point, one may be interesting in select among all the self-adjoint operators, the ones that represent reasonable physical quantities. This can be done using group theory. In particular, the symmetry group of a non-relativistic physical theory is the Galilean group. Hence to obtain interesting physical quantities, one have to represent this group over this mathematical structure. In order to do that, one have to specify on what Hilbert space we want to represent the group. The usual choice is the following
	\begin{center}
		\emph{The Hilbert space describing a single quantum particle in $\Rea^n$ with $m$ internal degree of freedom is $\Hi_{1p} = L_2(\Rea^n,dx)\otimes \Comp^m$.}
	\end{center}
	When one try to represent the Galilean group over the Hilbert space selected above demanding that the transition probabilities are preserved (i.e. looking for unitary representation), one is lead to consider the central extension of this group. Then, by the \emph{Stone-von Neumann-Mackey theorem} one can prove that the position and the momentum operator take its usual form, in addition from the \emph{Stone theorem} one can obtain the usual (free) hamiltonian operator. All these fact are contained in the following request
	\begin{center}
		\emph{The symmetry group of a non-relativistic quantum particle is the (central extension) of the Galilean group.}
	\end{center}
	Finally one need to explain how to deal with composite systems, namely system where there are more particles. Requiring that the structure of the single quantum particle systems is preserved, the measurement on one system does not disturb the other and that the maximal information content is constant irrespective to the way in which we gain the information, one is lead to the notion of tensor product of Hilbert spaces.
	\begin{center}
		\emph{The Hilbert space for a quantum system composed by $N$ quantum particles is given by $\Hi_{Np} = \bigotimes_{i=1}^N \Hi_{i, 1p}$.}
	\end{center}
	Once we have this method to treat composite system, then it is very reasonable to think that the only propositions that make sense are the one that does not depend on the ordering of the subsystems, namely the one that are invariant under the permutation of the single subsystems. Thus we are lead to
	\begin{center}
		\emph{For a quantum system which consist in $n$ subsystems, the physically admissible proposition are all the propositions that are invariant under the action of the permutation group of $\{1,\cdots,n\}$.}
	\end{center}
	Till now all the rules sketched to describe a quantum system, all consequence of the assumption that we have to use an Hilbert space to describe a quantum system and on that the invariance under a symmetry group is a reasonable physical requirement. Nevertheless, to conclude this discussion one need to introduce a last rule: the so called \emph{measurement postulate}.
	\begin{center}
		\emph{If at the time $t$ we find that a particular proposition $A$ holds, the state right after the measurement is given by
			\begin{equation*}
				\hat{\rho}' = \frac{\hat{P}_A\hat{\rho}\hat{P}_A}{\Tr{\hat{P}_A\hat{\rho}}}
			\end{equation*}
			where $\hat{P}_A$ is the projector associated to the proposition $A$.}
	\end{center}  
	Different proposal have been done to derive this postulate from the others like decoherence, dynamical collapses models and quantum bayesianism.

	\section{Modelling a lab: operational reconstruction of QM}\label{sec3}

	In this section we will briefly review the algebraic formulation of quantum mechanics, derived from operation considerations. The operational approach for the construction of a physical theory can be summarised as the attempt to formulate a physical theory defining each abstract mathematical operation as a procedure that can be executed, at least in principle, in a laboratory. This approach will lead to a formulation of quantum mechanics that is (almost) equivalent to the one described in the previous section, but starting from different assumptions. We will not speak anymore about propositions but we will focus our attention on physical quantities we use to describe a system. The main reference for this section are \cite{fS} for the operational arguments, \cite{vM} for the mathematical theorems and \cite{lA} for the probabilistic notions.
	
	\subsection{A bit of math: some notions of algebra}
	
	In this paragraph we will concentrate mostly of the mathematical notions and simple results we will use in what follows. The key concept is
	\begin{definition}
		An \emph{associative algebra} $\A$ over a field $\field$, is a $\field$-vector space equipped with a product $\star:\A \times \A\rightarrow\A$ such that
		\begin{enumerate}
			\item[1.] $a \star (b \star c) = (a \star b) \star c$, $\forall a,b,c \in \A$;
			\item[2.] $a \star ( b + c) = a\star b + a\star c $, $\forall a,b,c \in \A$;
			\item[3.] $(a+b) \star  c = a\star c + b\star c $, $\forall a,b,c \in \A$;
			\item[4.] $\alpha (a \star b) = (\alpha a)\star b = a \star (\alpha b)$, $\forall \alpha \in \field$ and $a,b \in \A$.
		\end{enumerate}
	\end{definition}
	Briefly, an algebra is a vector space equipped with a product operation which is associative and distributive with respect the vector space operations.  Algebras can be classified according with the following definitions
	\begin{definition}
		Given an associative algebra $\A$, then
		\begin{enumerate}
			\item[1.] is said \emph{normed algebra}, if equipped with a norm $\| \cdot \|$ such that $\|a \star b\| \leqslant \| a\| \|b\|$, $\forall a,b \in \A$;
			\item[2.] is said \emph{banach algebra}, if normed  and at the same time $\A$ is a Banach space under the norm;
			\item[3.] is said \emph{$^*$- algebra}, if equipped with an involution $^*: a\mapsto a^*$, $\forall a \in A$;
			\item[4.] is said \emph{$C^*$-algebra}, if banach and $\|a^*\star a\| = \|a\|^2$ (said \emph{$C^*$-property});
			\item[5.] is said \emph{algebra with unit}, if there exist an element $\unit \in \A$ such that $a = a\star\unit =\unit \star a$;
			\item[6.] is said \emph{abelian (or commutative) algebra}, if $[a,b] = a \star b - b\star a = 0$, $\forall a,b \in \A$.
		\end{enumerate}
	\end{definition}
	The symbol $\star$ for the algebraic product will be omitted, if there are no ambiguities with the usual product, and we also set $\field = \Comp$. Another important  algebra, which is not in general associative, is the \emph{Jordan algebra}
	\begin{definition}
		A \emph{jordan algebra} $\A$ is a vector space, equipped with a bilinear form $\circ: \A \times \A \rightarrow \A$ such that $a \circ b = b \circ a$ and $a \circ (b \circ (a \circ a)) = (a \circ b) \circ (a \circ a)$ (\emph{jordan identity}).
	\end{definition}
	In what follow we will consider an associative algebra, in particular a $^*$-algebra. Its elements may be classified in similar manner of what is typically done for operators: $a$ is said  \emph{normal} if $a a^* = a^* a$, and \emph{self-adjoint} if $a = a^*$. Again, following the similarity with operators (which in may cases form an algebra) one can define also a notion of \emph{spectrum} in algebraic contest.
	\begin{definition}
		Given a Banach algebra with unit, $\A$, and consider $a \in \A$. The \emph{spectrum} of $a \in \A$ is the set defined as
		\begin{equation*}
			\sigma(a) := \{ \xi \in \Comp | \nexists(a - \xi \unit)^{-1} \in \A \}
		\end{equation*}
	\end{definition}
	The following result about the spectrum of a $C^*$-algebra is interesting for our discussion
	\begin{proposition}
		Let $a \in \A$ be a self-adjoint element of a $C^*$-algebra with unit, the $\sigma(a) \subset [-\|a\|,\|a\|] \subset \Rea$.
	\end{proposition}
	In addition, the notion of spectrum allows us to introduce a further classification between the elements of a Banach algebra with unit:
	\begin{definition}
		Given a Banach algebra with unit $\A$, an element $a \in \A$ is said \emph{positive} if self-adjoint and its spectrum is positive $\sigma(a) \subset \Rea^+$. The set of all the positive elements will be denoted by $\A^+$.
	\end{definition}
	Positivity of the elements allows us to define the positivity of linear functionals over the algebra. Such functionals play a special role in the algebraic formulation of quantum mechanics since they are interpreted as \emph{states}.
	\begin{definition}
		Let $\A$ be a Banach algebra with unit, a functional $\omega: \A \rightarrow \Comp$ such that
		\begin{enumerate}
			\item[1.] it is \emph{positive}, namely $\omega(a)\geqslant 0$, $\forall a \in \A^+$;
			\item[2.] is \emph{normalised}, namely $\omega(\unit) = 1$;
		\end{enumerate}
		is called \emph{state}.
	\end{definition}
	It can be proved, that over a $C^*$-algebra, a linear functional is positive if and only if it is bounded (in particular $\|\omega\| = \omega(\unit)$). This implies that over a $C^*$-algebra, a state is a continuous (since bounded) functional. Finally the following result about states is interesting for our discussion
	\begin{proposition}
		Let $\A$ be a $C^*$-algebra, take $a \in \A$ and consider $\alpha \in \sigma(a)$. Then there exist a unique state $\omega_\alpha : \A \rightarrow \Comp$, such that $\omega_\alpha(a) = \alpha$.
	\end{proposition}
	
	\subsection{Operational approach for a physical theory: description of a single physical quantity}
	
	Any experimental science is based on the reproducibility of experiments. One prepare the system in a certain \emph{configuration} and then perform a series of measurements on some \emph{physical quantity} from which one can prove or disprove a fact. Physics was the first science where this procedure was applied. The theoretical models we use to describe a physical system should be based, as much as possible, on the way one have access to the information that we learn in a measurement. This is the heart of the operational approach for the construction of a physical theory.
	
	Let $\mathcal{C}$ be the set of all the possible configurations in which a system can be prepared, and $\A$ be the set of all the physical quantities of the system we can measure. Consider a physical quantity $a \in \A$, the outcome of a measurement for a system prepared in the configuration $\omega \in \mathcal{C}$, will be labeled by $m_\omega (a)$, and can be operationally defined as the value that the pointer of the measuring device assume when we measure the quantity $a$. The \emph{result of a measurement} of the physical quantity $a \in \A$ for a system prepared in the configuration $\omega \in \mathcal{C}$, labeled $\omega(a)$, is defined to be the average of all the measured value $m_\omega(a)$, repeating the experiment (ideally) an infinite number of times, namely
	\begin{equation*}
		\omega(a) := \lim_{n \rightarrow +\infty} \frac{1}{n} \sum_{i=1}^{n} m_{\omega}^{(i)}(a)
	\end{equation*}
	Two configurations $\omega_1, \omega_2 \in \mathcal{C}$ are \emph{operationally indistinguishable} if the result of the two measurements is the same for all the physical quantities. Mathematically speaking, on $\mathcal{C}$ we can say that $\omega_1 = \omega_2$ if and only if $\omega_1(a) =\omega_2(a)$, $\forall a \in \A$. Similarly, two physical quantities are operationally indistinguishable if preparing the system in all the possible configurations, the results of the measurements are always identical. Hence on $\A$ we can say that $a_1 = a_2$ if and only if $\omega(a_1) = \omega(a_2)$, $\forall \omega \in \mathcal{C}$. Using these last equivalence relation, we can define operationally, the usual mathematical operations over $\mathcal{C}$ and on $\A$. For example, consider $a \in \A$ and $\lambda \in \Rea$, one can define the physical quantity $\lambda a$ as the physical quantity measured by a measuring device whose pointer scale is dilated by a factor $\lambda$ with respect to the original scale. More formally, one can write $m_\omega(\lambda a) := \lambda m_\omega (a)$ $\forall \omega \in \mathcal{C}$, which implies that $\omega(\lambda a) = \lambda \omega(a)$. In similar way, given $a,b \in \A$, one can operationally define the sum of two physical quantities $a+b$ simply setting $m_\omega (a + b) := m_\omega (a) + m_\omega (b)$ $\forall \omega \in \mathcal{C}$, and so $\omega(a+b) = \omega(a) + \omega(b)$. We can see that, by definition, $\omega$ is \emph{linear}.
	
	One can operationally define also $a^n$, $n \in \Nat$ setting $m_\omega(a^n):=[m_\omega(a)]^n$ $\forall \omega \in \mathcal{C}$. Because of this definition, one can see that $m_\omega(a^0) = 1$ for any $\omega$. Thus we can define an \emph{identity} $\unit := a^0$. We can also see that it is quite reasonable to assume $a^{n+m} = a^na^m$, since $m_\omega(a^{n+m}) = m_\omega(a)^n m_\omega(a)^m$. Finally, it is not difficult to see that even for $\lambda \in \Comp$, $\lambda a$ is well defined if we measure separately the real and imaginary part of it. Hence, all the complex polynomials of $a \in \A$, like $\alpha_n a^n + \alpha_{n-1} a^{n-1} + \cdots + \alpha_0 \unit$, are well defined from the operational point of view. Let $\A_a$ denote the set of all the possible complex polynomials of $a$.  Clearly $\unit \in \A_a$ and on $\A_a$ one can naturally define an \emph{involution} setting $(\lambda a)^* := \overline{\lambda} a$, with $\overline{\lambda} $ complex conjugate of $\lambda $, and $(ab)^* = b^*a^*$. Thus we can conclude that \emph{$\A_a$ is an abelian $^*$-algebra with unit}. We cannot say the same thing on $\A$, because we may have troubles in the definition of $m_\omega(ab)$ and $m_\omega(ba)$, because of the property \textbf{O2}: for this reason we concentrate on $\A_a$ only.
	From the discussion done till now, we may conclude that the following things holds for any $b,c \in \A_a$ (hence polynomial of $a$)
	\begin{enumerate}
		\item[1.] $m_\omega(\alpha b + c) = \alpha m_\omega(b) + m_\omega(c)$;
		\item[2.] $m_\omega (bc) = m_\omega(b) m_\omega(c)$;
		\item[3.] $m_\omega (\unit) = 1$;
		\item[4.] $m_\omega(b^*) = \overline{[m_\omega(b)]}$;
		\item[5.] $m_\omega (b^*b) \geqslant 0$.
	\end{enumerate}
	We observe that the point 2, since we are dealing with polynomial of $a$ only, is a consequence of 1 and the properties of powers for the elements  of $\A_a$ and does violate \textbf{O2}. These have consequences on the configurations
	\begin{enumerate}
		\item[1.] $\omega(\alpha a + b) = \alpha \omega(a) + \omega(b)$;
		\item[2.] $\omega(\unit) = 1$;
		\item[3.] $\omega(a^*) = \overline{[\omega(a)]}$;
		\item[4.] $\omega(a^*a) \geqslant 0$
	\end{enumerate}
	This means that a configuration $\omega: \A_a \rightarrow \Comp$ is a positive, normalised linear functional over $\A_a$: for this reason $\omega$ is a state on $\A_a$.  At this point it is useful to introduce the following map for any $a \in \A_a$
	\begin{equation*}
		\| \cdot \|: \A_a \rightarrow \Rea \qquad \qquad \|a\|:= \sup_{\omega \in \mathcal{C}}|\omega(a)|
	\end{equation*}
	This number is the maximum value that a physical quantity may assume. Because any real instrument have a finite scale, $\|a\|$ is a \emph{finite}, positive real number. Without proving, we state the following proposition containing all the key properties of $\|\cdot\|$.
	\begin{proposition}
		Take $a \in \A$, $\lambda \in \Comp$ and $b,c \in \A_a$. The map $\|\cdot\|$ on $\A_a$ defined as above fulfils
		\begin{enumerate}
			\item[1.] $\|b+c\|\leqslant\|b\| + \|c\|$ and $\|\lambda b\| = |\lambda| \|b\|$;
			\item[2.] $\|b^*\| = \|b\|$;
			\item[3.] $\| bb^*\| \leqslant \|b\|^2$ with equality if and only if $b =b^*$;
			\item[4.] $\|bc\| \leqslant \|b\| \|c\|$ when $b=b^*$ and $c=c^*$.
		\end{enumerate}
	\end{proposition}
	Now, define following symmetric product on $\A_a$
	\begin{equation*}
		b \circ c := \frac{bc + cb}{2} \qquad\qquad b,c \in \A_a
	\end{equation*}
	It is not difficult to see that it is commutative, in addition also the Jordan identity is fulfilled. Thus $\A_a$ equipped with this product is a \emph{Jordan algebra}. Now, because $|\omega(bc)| \leqslant |\omega(b)||\omega(c)|$ on $\A_a$, one can conclude that $\|b \circ c\|\leqslant\|b\| \|c\|$ (and so $\|\cdot\|$ defines a norm), and so $\A_a$ is also normed, and after completion, $\A_a$ can be considered as a \emph{abelian Jordan-Banach algebra} (with unit). From the definition given for the $\circ$ product, one can conclude that $\A_a$ is is a sub-algebra of a larger abelian $C^*$-algebra with unit\cite{fS}. Thus, using operational arguments, we arrived to the following conclusion: each physical quantity can be described using an abelian $C^*$-algebra with unit. In the next paragraphs, we will see how it is possible to describe algebraically the whole physical system at the same time and not single physical quantities.
	
	\subsection{Another bit of math: represent an algebra}
	
	In the previous paragraph, we have seen that if when we want to describe a physical quantity, it is reasonable to use a $C^*$-algebra. Nevertheless, to explicitly perform any calculation, it should be useful to have more simple (less abstract) mathematical objects. In this section, we will briefly review how to represent an abstract $C^*$-algebra using concrete mathematical objects, like functions or operators.
	
	The first result in this sense, for the particular case of a commutative $C^*$algebra, is the so called \emph{commutative Gelfand-Naimark theorem}. To formulate it we need the following definitions
	\begin{definition}
		Let $\A$ be a commutative $C^*$-algebra with unit. A \emph{character of $\A$} is a non-zero homomorphism between $A$ and $\Comp$, namely a map $\phi:\A\rightarrow \Comp$ such that $\phi(ab) = \phi(a)\phi(b)$, $a,b \in \A$. The set of all characters of $\A$ is called \emph{structure space} (or \emph{spectrum of the algebra}), and is labeled by $\Delta(\A)$.
	\end{definition}
	Now, define a maps $\tilde{a}:\Delta(\A) \rightarrow \Comp$ as $\tilde{a}(\phi) := \phi(a)$, for $a \in \A$. This map is the so called \emph{Gelfand's transform} and can be proved that
	\begin{equation*}
		\sigma(a) = \{ \tilde{a}(\phi) \st \phi \in \Delta(\A) \}
	\end{equation*}
	which is the link between the spectrum of an element of the algebra $\sigma(a)$ and the structure space $\Delta(\A)$. At this point we can state the aforementioned theorem
	\begin{theorem}[commutative Gelfand-Naimark theorem]
		Let $\A$ be a commutative $C^*$-algebra with unit, and consider the algebra of continuous functions of the structure space $C(\Delta(\A))$ ($C^*$-algebra with respect to the norm $\|\cdot \|_{\infty}$). Then the Gelfand's transform is an isomorphism preserving the involution and the norm (isometric $^*$-isomorphism) between $\A$ and $C(\Delta(\A))$
	\end{theorem}
	Thus this theorem guarantee that we can always \textgravedbl represent \textacutedbl a commutative $C^*$-algebra with an algebra of continuous functions. As we will see this is no longer true if we drop the commutativity of the algebra: in this case we have to represent the algebra in a different manner.
	
	Let us now consider the general case of a $C^*$-algebra, without assume commutativity. It turns out that Hilbert spaces are the suitable concrete mathematical object on which we can represent this algebra. This motivate the following
	\begin{definition}
		Consider a $C^*$-algebra with unit $\A$ and an Hilbert space $\Hi$. An homomorphism, preserving the involution and unit, $\pi: \A \rightarrow \boundo$, is called \emph{representation of $\A$ on $\Hi$}. The representation $\pi$ is said \emph{faithful} if it is one-to-one. Finally, a vector $\psi \in \Hi$ is said \emph{cyclic} for $\pi$, if $\overline{\{ \pi(a)\psi \st a \in \A \}} = \Hi$.
	\end{definition}
	At this point, take the $C^*$-algebra with unit $\A$, and a state $\omega: \A \rightarrow \Comp$ on it. It is not difficult to see that $\langle a, b \rangle_\omega := \omega(a^*b)$ define pre-inner product (in general it is degenerate). In order to have a well defined inner product, one can define the set
	\begin{equation*}
		N_\omega := \{ a \in \A \st \omega(a^*b) = 0, \forall b \in \A \}
	\end{equation*}	
	Then on $\A/N_\omega$, the product $\langle \cdot , \cdot \rangle_\omega$ is a well defined inner product. Thus, once can complete $\A/N_\omega$ to an Hilbert space, and we will label such Hilbert space by $\Hi_\omega$. One may define for each $a \in \A$
	\begin{equation*}
		[\pi_\omega (a)] ([b]) := [a][b] \qquad\qquad \forall [b] \in \Hi_\omega 
	\end{equation*}
	This is by construction an homomorphism preserving the involution. Computing the norm, one can prove that $\pi_\omega(a)$ is bounded, thus $\pi_\omega$ is a representation of $\A$ on $\Hi_\omega$. Finally, we can understand the the unit of the algebra, defines a special vector in $\Hi_\omega$, which is $\Psi_\omega = [\unit]$. Such vector is cyclic for $\pi_\omega$. Thus we can see that, given a $C^*$-algebra $\A$ and a state $\omega$, we can always construct a triple $(\Hi_\omega, \pi_\omega, \Psi _\omega)$, called \emph{GNS triple}. The discussion done above, sketched a part of the proof of the so called \emph{Gelfand-Naimark-Segal theorem}.
	\begin{theorem}[GNS theorem]
		Let $\A$ be a $C^*$-algebra with unit and $\omega:\A\rightarrow\Comp$ a state. Then
		\begin{enumerate}
			\item[i)] there exist a triple $(\Ho,\pi_\omega,\Psi_\omega)$ where $\Ho$ is an Hilbert space, $\pi_\omega : \A \rightarrow 
			\boundoo$ is a $^*$-representation  of $\A$ on the $C^*$-algebra of bounded operators on $\Ho$, and 
			$\Psi_\omega \in \Ho$ is a vector, such that
			\begin{enumerate}
				\item[a)] $\Psi_\omega$ is unit vector, cyclic for $\pi_\omega$.
				\item[b)] $\langle \Psi_\omega | \pi_\omega(a) \Psi_\omega \rangle = \omega(a)$ for any $a \in \A$.
			\end{enumerate}
			\item[ii)] If $(\mathcal{H},\pi,\Psi)$ is a triple such that
			\begin{enumerate}
				\item[a)] $\mathcal{H}$ is an Hilbert space, $\pi: \A \rightarrow \boundo$ is a $^*$-representation and
				$\Psi \in \mathcal{H}$ is a unit vector cyclic for $\pi$;
				\item[b)] $\omega(a) = \langle \Psi | \pi(a) \Psi \rangle$;
			\end{enumerate}
			then there exit a unitary operator $\hat{U}:\mathcal{H}\rightarrow\Ho$ such that $\Psi = \hat{U}\Psi_\omega$ and 
			$\pi_\omega(a) = \hat{U}\pi(a)\hat{U}^{-1}$ for any $a \in \A$.
		\end{enumerate}
	\end{theorem}
	Thus we have found a way to \textgravedbl represent\textacutedbl a generic $C^*$-algebra over an Hilbert space. Nevertheless, such representation depend on the state, hence changing the state we change also the operators. In addition in general is not one-to-one: we may find element in $\boundoo$ which doesn't represent any element of $\A$. The way out of this problem is the \emph{Gelfand-Naimark theorem}
	\begin{theorem}
		Any $C^*$-algebra with unit $\A$ is isomorphic (there exist an isomorphism preserving the involution) to a sub-algebra of $\boundo$ for some Hilbert space $\Hi$.
	\end{theorem}
	In particular, one can prove that the Hilbert space considered in this theorem is $\Hi = \bigoplus_{\omega \in \mathcal{C}} \Hi_\omega$ and the isomorphism is ${\Pi(a):= \bigoplus_{\omega \in \mathcal{C}}\pi_\omega(a)}$. 
	
	Summarising we have that, any $C^*$-algebra with unit may be represented over an Hilbert space using suitable bounded operators, if the algebra is also abelian we can always find an algebra of function isomorphic to the original algebra.
	
	\subsection{Operational approach for a physical theory: description of a physical system}
	
	We have seen that, when we try to describe a single physical quantity, we are naturally lead to introduce an abelian $C^*$-algebra with unit. Commutativity, allows us to say that we can represent the elements of an algebra using functions, while the \emph{Riesz representation theorem} allows us to understand how to represent algebraic state over this space of function
	\begin{theorem}[Riesz representation theorem]
		Let $\X$ be a locally compact Hausdorf space and $\omega: C(\X) \rightarrow \Comp$ a continuous functional. Then there exist a unique Borel measure $\mu_\omega$
		such that
		\begin{equation*}
			\omega(f) = \int_{\X} f(x) \mu_{\omega}(dx)
		\end{equation*}
	\end{theorem}
	This theorem allows us to conclude that, given the couple $(\A,\omega)$, where $\A$ is a commutative $C^*$-algebra and $\omega$ is a state, then there exist a probability space $(\X,\borelx, \mu_\omega)$, and the elements of the algebra are continuous function on $\X$ (which means that they are random variables). Thus, using this different approach, one obtain that each physical quantity is always described by a random variable over a probability space. This is part of the content of the observation \textbf{O1} about a quantum system: we can see that it is a general feature of any physical system, since no quantum assumption was done in its derivation.
	
	From the discussion done above, one can see that $(\A,\omega)$, and the random variables over $(\X,\borelx,\mu_\omega)$ describe the same thing: thus one can choose if deal with probability using measure-theoretic notions or algebraic notions. More formally \cite{lA}, the couple $(\A,\omega)$ where $\A$ is a $^*$-algebra and $\omega$ is a positive normalised functional on $\A$, is called \emph{algebraic probability space}. If the algebra is $C^*$, $(\A,\omega)$ is said \emph{$C^*$-probability space}. Finally, it is possible to prove that any commutative algebraic probability space is equivalent, up to zero-measure sets, to the usual measure-theoretic probability space \cite{hM}.
	
	In general a physical system is described by more than one physical quantity. The mathematical object that we need to describe the whole systems should contains all the abelian $C^*$-algebra of all the physical quantities. Thus, in order to be conservative as much is possible, we may associate to each physical system, a $C^*$-algebra $\A$
	defined to be the smallest $C^*$-algebra which contains all the abelian $C^*$-algebra with unit associated to the single physical quantity $\A_a$ (clearly $\A$ will have unit). In addition we may extend by continuity all the states $\omega: \A_a \rightarrow \Comp$, on $\A$. Thus we can summarise as follow: \emph{to each physical system we may associate a $C^*$-probability space $(\A,\omega)$}. This in this way, one can motivate operationally , the observation \textbf{O1}, seen in the introduction.
	As we will see this $C^*$-probability space is not abelian in general. Given a $C^*$-algebra $\A$, for each $a \in \A$ and $\alpha \in \sigma(a)$, we may find a state $\omega_\alpha(a) = \alpha$. Thus we can see that the spectrum of an element of the algebra can be interpreted as the set of all the possible values that the physical quantity may assume (which is not the set of all the possible outcome of an experiment). Because the measurable quantities are always expressed using real numbers, we have to conclude that the spectrum is a subset of $\Rea$. This happens if and only if the physical quantities are represented by self-adjoint elements of $\A$. Thus \emph{physical quantities are the self-adjoint elements of the algebra}.
	
	At this point we want to analyse more in detail the structure of the $C^*$-algebra $\A$. In particular we want to discuss under which condition it is abelian. The main tool we will use is the notion of \emph{entropy of a physical quantity}. Let $a \in \A$ be a physical quantity and $\omega$ a state on it. By the GNS theorem, we may always find a triple $(\Ho,\pi_\omega, \Psi_\omega)$, and so the physical quantity $a \in \A$, can be  represented on $\Ho$ as $\hat{\pi}_\omega(a)$. Using the spectral measure of this operator, one can compute the probability to observe the value of $a$ in the set $A$ as
	\begin{equation*}
		P_\omega(a \in A) = \int_{\sigma(a)} \chi_A(x) \mu^{(\hat{\pi}_\omega(a))}_\omega(dx) = \int_{[-\|a\|,\|a\|]} \chi_A(x) \mu^{(\hat{\pi}_\omega(a))}_\omega(dx)
	\end{equation*}
	since $\sigma(a) \subset [-\|a\|,\|a\|]$ and the spectral measure has support on $\sigma(a)$ only. Now, if we partition the spectrum using sets of diameter $\epsilon$, namely $\cup_{i \in I} U_i^{(\epsilon)} = [-\|a\|,\|a\|]$, for a given $\epsilon$ we obtain a set of probabilities $\{ p_\epsilon (i;\omega)\}_{i \in I}$, where $p_\epsilon (i;\omega):= P_\omega(a \in U_i^{\epsilon})$, which describe the probabilistic behaviour of the physical quantity in the state $\omega$. The diameter $\epsilon$ of the above sets, can be thought as the width of the bins of the measuring device of the given physical quantity. Then at this point one can define consistently the \emph{$\epsilon$-entropy of $a$ in the state $\omega$} as
	\begin{equation*}
		H_\omega^{(\epsilon)}(a) := - \sum_{i \in I} p_\epsilon(i;\omega) \log p_\epsilon(i;\omega)
	\end{equation*}
	This is the entropy of the physical quantity $a \in \A$ for a system prepared in the state (configuration) $\omega$, when is measured with an measurement apparatus with resolution $\epsilon$. Given a second physical quantity $b \in \A$, in the same way, one can define $H_\omega^{(\delta)}(b)$. Then, excluding the trivial partitions, the following theorem holds
	\begin{theorem}\label{myteo}
		Let $\A$ be a $C^*$-algebra with unit and consider $a,b \in \A$. If for any state $\omega$, and any $\epsilon,\delta > 0$
		\begin{equation*}
			H_\omega ^{(\epsilon)}(a) + H_\omega^{(\delta)}(b) \geqslant D(\epsilon, \delta)
		\end{equation*} 
		with $ D(\epsilon, \delta)$ positive and fixed for any $\epsilon$ and $\delta$ positive, then $[a,b] \neq 0$.
	\end{theorem}
	Also the converse holds \cite{MU}, thus the above relation is another way to demand for non-commutativity between $a$ and $b$. Operationally, the meaning is simple: even if we know $a$ with certainty ($H_\omega ^{(\epsilon)}(a) = 0$) then we can't know $b$ with arbitrary precision, and viceversa. This the operational analogous of the observation \textbf{O2}. The above relations involve only probability, and so it can be considered as an operational criterium for establish if two physical quantities are represented by commuting element of $\A$ or not. Hence, as claimed above, in general $\A$ may not be commutative. A final observation: the partition of the spectrum is necessary if we admit the possibility to have observable with $\sigma(a)$ which is not discrete (algebraically this cannot be established in advance, because the usual classification of the spectrum depend on the state), necessary to have a well defined notion of (Shannon) entropy. If one deal with physical quantity having $\sigma(a)$ discrete (for instance by construction), the theorem can be stated using only the usual Shannon's entropy without consider any partition of $\sigma(a)$. Relations like the one described above, are called \emph{entropic uncertainty relations}. Hence we can see that, \emph{if we find an entropic uncertainty relation, then the algebra cannot be abelian}. In this case, such algebra can be represented only using a subset of bounded operators over an Hilbert space $\Hi$, and not with functions.  
	
	\subsection{Algebraic formulation of quantum mechanics}
	
	Now, we are able to reconstruct quantum mechanics from the operational point of view. Below we will list the set of axioms needed to reconstruct quantum mechanics, justified by the discussion done till now.
	\begin{center}
		\emph{Each physical system is described by an $C^*$-probability space $(\A,\omega)$. The physical quantities are the self-adjoint elements of $\A$ and the possible way one can prepare the system are represented by states $\omega$.}
	\end{center}
	This is a general feature of a physical system and no quantum assumption are done till this point. The probabilistic interpretation, allows us to conclude that
	\begin{center}
		\emph{For a composite system, the algebraic probability space is constructed using the $C^*$-tensor product of the algebras associated to each subsystem.}
	\end{center}
	The $C^*$-tensor product, is the only possible tensor product between algebras which preserves the $C^*$-property. This rules is just the generalisation in $C^*$-algebraic contest of what is usually done in measure-theoretic probability spaces: in fact, for commutative $C^*$-algebra, it reduces exactly to the cartesian product of the sample spaces. One need to specify if the algebra is abelian or not.
	\begin{center}
		\emph{For a quantum system, some entropic uncertainty relation holds, which means that $\A$ is not abelian.}
	\end{center}
	This requirement force us to remain over an Hilbert space, when we represent the $C^*$-algebra. Positive, normalised linear functional are states and the GNS theorem show that the expectation is computed as in the previous section. Representing the algebra $\A$ over an Hilbert space, one is forced to consider physical quantities only some class bounded operators. Thus, we have no room for the position and momentum operators, because they are unbounded. Nevertheless there is a way out: one can always consider succession of bounded operators converging to the unbounded position and momentum operators. The commutation relations between them can be obtained by Weyl relations defining a \emph{Weyl $C^*$-algebra}. Time evolution can be obtained using the additive group $(\Rea,+)$, which is the same subgroup of the Galilean group generating the evolution in the previous section, to describe time translations.
	\begin{center}
		\emph{Time evolution is described by the group of time translations $(\Rea,+)$.}
	\end{center}
	On the GNS Hilbert space, such group is represented via Stone theorem, obtaining the usual Schrodinger evolution. Finally one need to introduce the measurement postulate.
	Using algebraic probabilistic considerations, the measurement postulate can be obtained using conditional expectation (this is the quantum bayesanisim solution of the measurement problem), nevertheless this solution is not universally accepted. Thus one should also need
	\begin{center}
		\emph{If the observed value of $a \in \A$ in the state $\omega$ is in the Borel set $F$, then the state after the measurement $\omega_F$ is}
		\begin{equation*}
			\omega_F : b \in \A \mapsto \frac{\langle \Psi_\omega | \hat{P}_F \pi_\omega(b) \hat{P}_F \Psi_\omega \rangle}{\langle \hat{P}_F\Psi_\omega | \hat{P}_F \Psi_\omega \rangle}
		\end{equation*}
	\end{center}

	\section{What is the meaning of an Hilbert space? A toy-model proposal}\label{sec4}

	In the two previous sections, we have seen in a rather informal way, how the quantum mechanics can be formulated starting from basic principles. In both cases, two are the basic assumptions: one about the intrinsic probabilistic nature of a physical system (a very general feature), and one about the limitations in this description due to the presence of quantities that cannot be known at the same time. We have also seen that is this feature that render unavoidable the Hilbert space formulation. In the quantum logic approach, is this feature (\textbf{O2}) which for us to abandon the usual (boolean) logic, and so the possibility to use the ordinary probability theory. In the operational-algebraic approach, are the entropic uncertainty relations that make the algebra not abelian, and so we cannot map it completely into an ordinary probability space.
	
	Quantum mechanics is usually formulated using the Hilbert space language, and this give rise to all the famous problems in the interpretation (like state-superposition or entanglement). We cannot abandon such formulation, because of non-commutativity, so the proposal of this section, is to treat non-commutativity as the reason because one is forced to use Hilbert space, and doesn't consider it as a consequence, as it is typically done. Once one accept this view, one can recognise that the entropic uncertainty relations and algebraic probability spaces are a very useful tools to derive the limitations in the simultaneous description of physical quantities from the intrinsic probabilistic nature of a physical system. In the following paragraph we will describe a simple toy-model, where the origin of the uncertainty is clear, which is able to reproduce the non-commutativity at algebraic level, of the analogous position and momentum for a particle. This model is very simple, and we do not claim that it reproduces completely the quantum analogous of it, nevertheless we think that it suggests an interesting motivation on the reason because nature at fundamental level, should be described using Hilbert spaces.
	
	\subsection{The toy model: random jumps over a random space}
	
	In the model proposed here, there are two main actors: the space and the particle. The space is assumed discrete and finite: it can be thought as a random distribution of, say $N$, points over the real line (we will consider the 1-dimensional case for simplicity). In addition the space is not assumed static but stochastic: the initial distribution of space points is assumed to bee know and it evolve in time as if each space point is a discrete-time random walk. The particle is assumed to be a point-like object whose dynamics can be thought as a succession of jumps from one space point to another. Now, we formalise this ideas from the mathematical point of view,
	
	Let us start briefly reviewing the mathematical formulation of the basic stochastic process describing the space: \emph{the random walk on a line}.	Let $\{X_i\}$ be a collection of random variables such that $P[X_i = +l] = p$ and $P[X_i = - l] = 1 - p =: q$, where $l$ is a fixed number. Such processes are called \emph{Bernulli random variables} and typical example of it is the coin tossing. We may describe the random walk in the following way. Suppose we have a person in the initial position $x_0$ at the initial (discrete) time $0$. At each instant of time this person tosses a coin: if he gets head he will move on the left of $l$, otherwise he will move on the right of the same amount. Thus the position of the person at the time $1$ will be $S_1 = x_0 + l$	if he get head, $S_1 = x_0 - l$ otherwise. Repeating this procedure for each instant of time we can say that the movement of this person is a random walk. Hence we can say that the	random walk starting at the point $x_0$, $S_N$, is defined as
	\begin{equation*}
		S_N := x_0 + \sum_{i=0}^N X_i
	\end{equation*}
	Central quantity we want to compute is the probability to have $S_N = d$, for some $d$ integer multiple of $l$, at the time $N$ starting from the point $x_0$. It can be proved that
	\begin{equation*}
		P[S_N - x_0 = d] =
		\begin{cases}
			\frac{N!}{\left[\frac{1}{2}\left(\frac{d}{l} + N\right)\right]!\left[\frac{1}{2}\left(N - \frac{d}{l}\right)\right]!} p ^{\left(\frac{d}{l} + N\right)}q^{\left(\frac{d}{l} + N\right)}&
			\mbox{  if  }d \in [-Nl,+Nl] \\
			0 &\mbox{  otherwise}
		\end{cases}
	\end{equation*}
	Suppose that the initial position $x_0$ is a random variable with distribution $\pi(x_0)$ over the real line (possibly discrete), then we can write that
	\begin{equation*}
		P[S_N = d] = \int_\Rea dx_0  P[S_N - x_0 = d]\pi(x_0)
	\end{equation*}
	Consider now a collection of random walks on the line, $\{S_N^{(i)}\}_{i \in I}$, starting in different points $x_{0}^{(i)}$, where $I$ is a finite set. At each instant of time, this collection of random walk will select at most $|I|$ points over the real line (two or more random walks may overlaps because they are assumed independent: the presence of a random walk in a give point doesn't influence the probability of another random walk to be found in the same point). Such collection of points, randomly distributed over the real line, is our model of space.
	
	The particle in this model can be described essentially by two quantities: the first is the position of the particle, $X_N$, the second its velocity, $V_N$, suitably defined over a space of this kind. Let us consider the position first. Let $\Pi_N$ be the finite set of points selected by the collection of random walks $\{S_N^{(i)}\}_{i \in I}$ at the time $N$. Clearly $\Pi_N \subset \Rea$. The random variable describing the position of a point-like particle at time $N$ will be a map $X_N: \Omega_{X_N} \rightarrow \Pi_N$. At this level seems problematic define the possible outcome of $X_N$ because the set $\Pi_N$ change at each instant of time. To avoid this problem, we may think that the random variable $X_N$ will select a single random walk in the collection $\{S_N^{(i)}\}_{i \in I}$. Select means that at the time step $N$, we have that $X_N = S_N^{(i)}$, in probability. Hence, in general we may relate the probability to observe the particle in a given point of space, with the probability to find a point of space where we observe the particle, as
	\begin{equation}\label{PXN}
		P[X_N = c] = \sum_{i \in I} P[X_N = S_N^{(i)}| S_N^{(i)} = c]P[S_N^{(i)}=c]
	\end{equation}
	Let us explain better the quantities involved in this equation.	We have
	\begin{enumerate}
		\item[a)]$P[X_N = S_N^{(i)}| S_N^{(i)} = c] =: \gamma(N,i,c)$ represent the probability that the particle select the $i$-th random walk in the collection, assumed that this 
		random walk at the time $N$ is in $x = c$. This is the probability that can be changed if we act on the particle only. More precisely, we can prepare the system (in our case the particle) in a configuration such that the probability to observe it in a given point is higher respect to another configuration, changing this term. For example, if we want to increase the probability to observe the particle in $x = a$, we may select (hence choose the index $i$) the random walks in the collection with an higher probability to be found in $x = a$. Summarising, when we prepare the particle in a given configuration we act on this object.
		\item[b)] $P[S_N^{(i)}=c]$ is the probability to observe a point of space in $x = a$. This quantity is given by the model and cannot be changed when we act on the particle
		only.
	\end{enumerate}
	The jumps of the particle between two different points of space will be modelled with a simple discrete-time Markov chain with transition probabilities
	\begin{equation}\label{forwardprobX}
		P[X_{N+1} = b | X_N = c] =: \alpha(b,c)
	\end{equation}
	where $b$ and $c$ are integer multiple of $l$, the step of the random walk (in what follow we will assume $l = 1$, for simplicity, and so $b,c \in \Nat$). These transition probabilities are assumed to fulfil some equation describing the physical system's dynamics. 
	As we will see, the random space described before will put some constraints on the possible values these transition probabilities may assume. Let us consider now the second random variable we are interested in: the velocity. The space described above is discrete and, in addition even the time is assumed discrete, hence it seem reasonable to define the velocity of a particle over this space as
	\begin{equation*}
		V_N := \frac{X_{N+1} - X_N}{N+1 - N} = X_{N+1} - X_N
	\end{equation*}
	Clearly, if it is a random variable it should depends both on $X_N$ and $X_{N+1}$. This fact render problematic the computation of the various probability generating function (like the characteristic functions) as a function of the random variable $X_N$ and $X_{N+1}$. Nevertheless, this problem may be bypassed, in a certain sense, following this intuitive idea. Suppose we know that the particle at the time $N$ is in the position $X_N = c$, then we know that the event $C:=\{X_N = c\}$ is true, namely $P[X_N = c] = 1$ (which means $P[X_N = d] = \delta_{d,c}$). Then, in this case, we can write that $V_N = X_{N+1} - c$ when $C$ is true. This suggests that the probability to have $V_N = a$ assumed that $C$ is true is equal to the probability that $X_{N+1} = a+c$ (where $a$ can vary and $c$ is fixed). Then, using (\ref{forwardprobX}) we can conclude that $P[X_{N+1} = a+c] = \alpha(a+c,c)$ and so $P[V_N = a|C] = \alpha(a+c,c)$. Finally we can write that $P[V_N = a]$ is given by
	\begin{equation*}
		\begin{split}
			P[V_N = a] &= \sum_c P[V_N = a | X_N = c]P[X_N = c]  \\
			&= \sum_c \alpha(a+c,c)P[X_N = c]
		\end{split}
	\end{equation*}
	This complete the probabilistic description of the particle in this model.
	
	\paragraph*{Remark.} The computation of $P[V_N = a | C]$ can be done in a rigorous way using characteristic function. Since $V_N = X_{N+1} - X_N$ and because when $C$ is true ($P[C] =1$, hence $P[X_N=d|C] = \delta_{d,c}$), $X_N$ and $X_{N+1}$ are independent, from the properties of the characteristic functions one can write $\varphi_{V_N}(\lambda)|_C = [\varphi_{X_{N+1}}(\lambda)|_C ][ \varphi_{ - X_N}(\lambda)|_C]$. This means that
	\begin{equation*}
		\begin{split}
			\varphi_{V_{N}} (\lambda)|_C &= \left(\sum_{b} P[X_{N+1}=b|C]e^{i\lambda b}\right)\left( \sum_{d} P[X_N = d|C] e^{i\lambda(-d)}\right) \\
			&= \sum_{b} P[X_{N+1}=b|C]e^{i\lambda (b-c)} \\
			&= \sum_{b} \alpha(b,c)e^{i\lambda(b-c)} 
		\end{split}
	\end{equation*}
	In the case of discrete random variable, one can recover the probability measure from the characteristic function using the formula
	\begin{equation*}
		P[X=a] = \lim_{T\rightarrow+\infty} \frac{1}{2T}\int^{+T}_{-T} e^{ita} \varphi_X(t)dt
	\end{equation*}
	Hence we have that
	\begin{equation*}\label{VNC}
		\begin{split}
			P[V_N = a|C] &= \lim_{T\rightarrow+\infty} \frac{1}{2T} \int_{-T}^{+T} e^{-i\lambda a}\varphi_{V_N}(\lambda)|_Cd\lambda \\
			&= \lim_{T\rightarrow+\infty} \frac{1}{2T} \int_{-T}^{+T} e^{-i\lambda a}\sum_b  \alpha(b,d)e^{i \lambda(b-c)}d\lambda \\
			&= \sum_b  \alpha(b,c)\lim_{T\rightarrow+\infty} \frac{1}{2T} \int_{-T}^{+T} e^{i\lambda(b-c-a)}d\lambda \\
			&= \sum_b  \alpha(b,c) \delta_{0,b-c-a} \\
			&= \alpha(a+c,c)
		\end{split}
	\end{equation*}
	Confirming the result obtained above.
	
	\subsection{Entropy for $V_N$ and entropy for $X_N$}
	
	In this paragraph, we will compute the entropy of the two random variables described before. Consider the following situation: suppose that we know that at time $N$ the particle is in the position $x=c$. Thus we can conclude that $P[X_N = d|C] = \delta_{d,c}$. In addition, using (\ref{VNC}), we also know that $P[V_N = a|C] = \alpha(a+c,c)$. Now, if $\alpha(a+c,c)$ can be changed continuously to a delta, then we should obtain $P[V_N = a|C] = \delta_{a+c,c}$. The computation of the two entropies with these probabilities will give us $H(X_N|C) = 0$ and $H(V_N|C) = 0$. Nevertheless we should observe the following fact. Recalling (\ref{PXN}), we can write 
	\begin{equation}\label{transprob}
		\alpha(a+c,c) = \sum_{j \in I} P[X_{N+1} = S^{(j)}_{N+1}|S^{(j)}_{N+1} = a+c, C] P[S^{(j)}_{N+1} = a+c]
	\end{equation}
	The quantity $P[X_{N+1} = S^{(j)}_{N+1}|S^{(j)}_{N+1} = a+c, C] =: \gamma(N+1,j,a+c|C)$ have the following property
	\begin{equation}\label{propG}
		\sum_{j \in I}\gamma(N+1,j,a+c|C) = 1
	\end{equation}
	Recalling that $ \gamma(N+1,j,a+c|C) \in [0,1]$ because is a transition probability, the equation (\ref{transprob}) can be considered as the average, with respect the probability distribution $\{ \gamma(N+1,j,a+c|C)\}_{j \in I}$, of the probability to find a point of space in $x = a+c$. This observation is very important: it allows us to find a bound for the transition probabilities. In fact, by the property of the average, for all the $\alpha(a+c,c) \neq 0$ (which gives a non zero contribution to the entropy) we can write that
	\begin{equation}
		\alpha(a+c,c) \leqslant \max_{j \in I} P[S^{(j)}_{N+1} = a+c]
	\end{equation}
	If we assume that the random space is a \emph{purely random process}, namely that $P[S^{(j)}_N = b] < 1$ for any $N$, $b$ and $j \in I$, then we can see that the $\alpha(a+c,c)$ cannot be $1$ for any value of $a$ (from now on it will be always assumed so). 
	This implies that we cannot have a delta-like probability distribution for both the observables $V_N$ and $X_N$ if the space is a purely random process. The bound found above cannot be changed if we act on the particle only: it is linked to the process describing the random space. This observation suggests that we can find a bound for the information that we can have on these two random variables at the same time.\newline
	
	Let us find a bound for the sum of the two entropies. It is a known fact that the entropy is a concave function of the probability distribution. Thus, the Jensen inequality holds, namely if $g$ is a concave function
	\begin{equation*}
		g\left( \frac{\sum_{i=1}^{n} a_i x_i}{\sum_{i=1}^{n} a_i} \right) \geqslant \frac{\sum_{i=1}^{n}a_i g(x_i)}{\sum_{i+1}^{n} a_i}
	\end{equation*}
	where $a_i \in \Rea$ are arbitrary numbers. Then, we can write
	\begin{equation*}\begin{split}
			H(V_N|C) & = -\sum_{a} \alpha(a+c,c) \log \alpha(a+c,c) \\
			& = - \sum_a \left( \sum_{j \in J} \gamma(N+1,j,a+c|C) P[S^{(j)}_{N+1} = a+c]\right) \log \left( \sum_{j \in J} \gamma(N+1,j,a+c|C)P[S^{(j)} = a+c] \right) \\
			& \geqslant - \sum_a  \sum_{j \in J} \gamma(N+1,j,a+c|C) \left( P[S^{(j)}_{N+1} = a+c] \log P[S^{(j)} = a+c] \right) \\
			& \geqslant - \sum_a \min_{j \in J}|_a \left( P[S^{(j)}_{N+1} = a+c] \log P[S^{(j)} = a+c] \right)
		\end{split}\end{equation*}
		Where we used the Jensen inequality and (\ref{propG}), while $\min_{j \in J}|_a$ is the minimum for $a$ fixed. This is already a bound on $H(V_N|C)$ which doesn't involve processes related to the particle, nevertheless we can further simplify this result if we add some reasonable assumption on the random walks describing the space process.
		We may require that
		\begin{enumerate}
			\item[a)] The initial positions $\{x^{(i)}_0\}_{i \in J}$ are $i.i.d.$ random variables, namely $x^{(i)}_0 \sim \pi^{(i)} = \pi$ for any $i \in J$, 
			\item[b)] The left and right probabilities (i.e. $p^{(i)}$ and  $q^{(i)}$) of the random walks are all equal, namely $p^{(i)} = p$ for any $i \in J$.
		\end{enumerate}
		Under these two assumption, we can say that all the random walks are \emph{statistically equivalent}, in the sense that $P[S_N^{(i)} = a] = P[S_N^{(j)} = a]$ for any $i,j \in J$ ("observing" the space process only at time-step $N$). This implies that
		\begin{equation*}
			- \sum_a \min_{j \in J}|_a \left( P[S^{(j)}_{N+1} = a+c] \log P[S^{(j)} = a+c] \right) = - \sum_a \left( P[S^{(j)}_{N+1} = a+c] \log P[S^{(j)} = a+c] \right)
		\end{equation*}
		namely
		\begin{equation*}
			H(V_N|C) \geqslant H(S_{N+1})
		\end{equation*}
		Finally, it is not difficult to see that $H(X_N|C) = 0$ since $P[X_N = d |C] = \delta_{d,c}$. Hence we can write that
		\begin{equation*}
			H(X_N|C) + H(V_N|C) \geqslant H(S_{N+1})
		\end{equation*}
		and because $H(X|Y) = \sum _i P[Y = i]H(X|\{Y=i\})$ and $H(X|Y) \leqslant H(X)$ (conditioning reduces entropy),  the above inequality implies that
		\begin{equation}\label{EUR}\begin{split}
				H(X_N) + H(V_N) &\geqslant \sum_c P(X_N = c)\left[H(X_N|C) + H(V_N|C) \right] \\
				&\geqslant \sum_c P[X_N = c]H(S_{N+1}) = H(S_{N+1})
			\end{split}\end{equation}
			The RHS has the following features
			\begin{enumerate}
				\item[a)] $H(S_{N+1})$ is a positive quantity which cannot be changed if we act on the particle only: $H(S_{N+1})$ is fixed once that the model of the space is given;
				\item[b)] $H(S_{N+1})$ is zero only for a deterministic space. This case is excluded if we assume that the random space is a purely random process (the space is not 
				deterministic). As expected from the initial discussion on the transition probabilities, we can see that the bound in the entropies is related to the random nature 
				of the space;
				\item[c)] It is not guarantee that this bound is optimal.
			\end{enumerate}
			The above bound can be explained in the following manner: we may change the system configuration (namely the $\{\gamma(N,i,c)\}$) in order to know completely the position of the particle, nevertheless the velocity of the particle remains uncertain at least as the future position of a space point.  Such uncertainty cannot be reduced acting on the particle only. The discussion done till now, should prove that  position and velocity of a point-like particle which jumps at random in different points over a random space satisfy an entropic uncertainty relation.
			
			\subsection{Algebraic description of a point-like particle}
			
			The final result of the previous paragraph, suggests that position and velocity doesn't commute. In order to point out explicitly this non-commutativity, we will describe the particle (not the space) using algebras. The position  is a random variable, namely a measurable map between a probability space $(\Omega_{X},\mathcal{E}(\Omega_{X}), P_{X})$ and a measure space, say $(\Rea,\borel)$. Thus we can write $X_N:\Omega_X \rightarrow \Rea$. Same things can be say for the velocity: it is defined over a probability space $(\Omega_V, \mathcal{E}(\Omega_V),P_V)$, take value over the measure space $(\Rea,\borel)$ and is a measurable map, thus $V_N: \Omega_V \rightarrow \Rea$. This description is equivalent, up to null sets, to the algebraic probability spaces $(\A_X, \varrho)$ and $(\A_V, \varsigma)$, for $X_N$ and $V_N$ respectively. In this description, $X_N$ and $V_N$ are algebraic random variable, namely (involution preserving) homomorphism between two algebras: $X_N$  is defined as the homomorphism $X_N:\A_X \rightarrow L_\infty(\Rea,\borel)_X$ and similarly for $V_N$ ($L_\infty(\Rea,\borel)$ is the canonical $C^*$-algebra associated to the measurable space $(\Rea,\borel)$).
			
			At this point, using the algebraic probability language, we can define the $C^*$-probability space describing the point-like particle of the model as $C^*$-probability space $(\A,\omega)$, where $\A$ is the smallest $C^*$-algebra which contain both $L_\infty(\Rea,\borel)_X$ and $L_\infty(\Rea,\borel)_V$, and $\omega$ is a state. Thus $X_N$ and $V_N$ are elements of $\A$, and the discussion done in the previous paragraph, tell us that they fulfil an entropic uncertainty relation and, by the theorem \ref{myteo} we have that $[X_N,V_N] \neq 0$. Thus, recalling the previous discussion about non-abelian algebras, we can see that the description of our point-like particle must be done over an Hilbert space, and cannot be mapped to an ordinary probability space. 
			
			It is worth to remark that this apparently strange result, is possible because we eliminated the space from the description. In fact, we start using an ordinary probability space, and we found a relations between the entropies of the three main objects of the model (the space, the position and the velocity). Then eliminating the space from the model, we found an entropic uncertainty relation. It is interesting to observe that in ordinary quantum mechanics the space doesn't play any role, exactly as in the final model of the particle.
			
			\subsection{Conclusions, weak points of the model and possible further development}
			
			The model presented in the previous sections have the following interesting feature: the space as well as the particle are treated in the same way. Thinking that any observation we can do in a laboratory give us only a probabilistic outcome (it is a nonsense to say "the quantity $A$ has the value $b$") this feature can be justified using an operational approach for the construction of a physical theory. In the model presented, the evolution of the particle is the only possible, in the sense that if the particle change position, it must happens with a jump. Thus in this sense we derived \textbf{O1} form \textbf{O2}, in the particular model considered. Of course, from the discussion done till now we cannot conclude that the quantum mechanics is equivalent to something similar to the model presented here. In particular, after a bit of though, the following aspects may look strange
			\begin{enumerate}
				\item[a)] \emph{Time is discrete}. The discreteness of time allows us to define $V_N$ as is done and non-commutativity seem to be a consequence of it. Nevertheless, in 	
				the proof of the entropic uncertainty relation between $X_N$ and $V_N$ time doesn't play any role. So the proof seem to be quite robust to possible change in the definition of $V_N$, due to change in the time assumptions.
				\item[b)] \emph{Space is discrete}. The structure of space is the core of the proof. Nevertheless, discreteness limit only the place where a point of space can be found.
				This is problem can be solved switching the description of space from random walks to wiener(-like) processes, something that at the moment is not available. Thus this weak point still remain but, we will discuss later an interesting mathematical object that can be used to treat this problem.
				\item[c)] \emph{The constant in the entropic uncertainty relation change with time}. Looking back to the entropic uncertainty relation (\ref{EUR}) one can easily see that 
				the constant depend on time. This seem rather strange despite it doesn't influence the result: the constant remains positive for each time at it is (particle) state independent. Nevertheless, one should observe this fact: the constant is related to the stochastic process describing the space. In particular, in (\ref{EUR}) only the RHS depend on the space, so if we change the model of space, only this side change. For example one can substitute the usual random walk (where the possible position of the walker can be any integer number, this cause the increasing of the value of the constant) with a reflexing boundary random walk. In this case the number of possible position of the point of space are a finite number, and so the constant can be fixed. This suggests that this problem is not so fundamental. In addition the bound is not assumed to be optimal, as already observed.
			\end{enumerate}
			From the above arguments, one can see that a better model for the space should be desirable. Keeping in mind this we may list a series of fact, suggesting that further studies in this direction can be interesting
			\begin{enumerate}
				\item[a)] \emph{Determinantal random point field.} A determinantal random point field is a stochastic process which describe the random distribution of $N$ (possibly 
				infinite) points over $\Rea^n$ (or more generally over a Polish space). This process is said deteriminantal because any point correlation function can be expressed as Fredoholm's determinant of a locally trace class operator $\hat{\sigma}$ over an Hilbert space (in particular $\hat{\sigma}:L_2(\Rea^n)\rightarrow \mathcal{K}$, with $\mathcal{K}\subset L_2(\Rea^n)$ and $\dim \mathcal{K} = N$) \cite{aS}. Such property is called \emph{determinantal property}. Locally trace class operator are operators whose trace may be infinite but, if restricted to suitable subspace, it is finite. Clearly any trace class operator can be considered as a locally trace class operator. Thus given a quantum system described by an Hilbert space $\Hi$ and a trace class operator $\hat{\rho}$ we may associate to it a determinantal random point field. Other similarity of these mathematical objects with quantum mechanics is that also that can be described using the second quantisation language. Finally  the unitary evolution of quantum mechanics preserves the determinantal property.
				\item[b)]\emph{$L_2(\Rea^3)$ for a particle.} If we assume that the space where the particles live is a random distribution of points over $\Rea^3$ described by a determinantal random point field, then we may justify the postulate of quantum mechanics which tells us that $L_2(\Rea^3)$ is the Hilbert space for a single particle.
				\item[c)]\emph{Position and momentum operator and Galilean group.} From the Stone-von Neumann theorem we may justify that the position and momentum operators in
				quantum mechanics are defined as
				\begin{equation*}
					\hat{X}_i \psi(x) = x_i \psi(x) \mspace{50mu} \hat{P}_i\psi(x) = - i\frac{\partial}{\partial x_i}\psi(x)
				\end{equation*}
				for $\psi(x) \in \mathcal{S}(\Rea^n)$.  In quantum mechanics doesn't seem a priori valid the classical relation between position and momentum for a point-like particle ($p = m\dot{x}$). Nevertheless from the Galilean group one can prove that\cite{vM}
				\begin{equation*}
					\langle P_t \rangle = m \frac{d}{dt} \langle X_t \rangle
				\end{equation*}
				Assuming, because a limited accuracy for any clock, that the time is discrete (hence the limit is replaced by an inferior) and using the Wigner quasi probability distribution, we can write that
				\begin{equation*}
					P_{t} = m \inf_{\delta t} \frac{X_{t + \delta t} -X_{t}}{\delta t} \mspace{50mu} dW\mbox{-a.s.}
				\end{equation*}
				Setting $\delta t =1$ ($\delta t$ is our unit of time) and $t = N\delta t$, we can write $P_N = m (X_{N+1} - X_N) = mV_N$. This consideration justify at qualitative level the interest in the two random variable analysed in these notes. 
			\end{enumerate} 
			We conclude this writing, with a quotation whose author consider it contains the spirit of what was written in this last section
			\begin{center}
				\textgravedbl \emph{Probability is the most important concept in modern science, especially as nobody has the slightest notion what it means} \textacutedbl
			\end{center}
			\begin{flushright}
				(Bertrand Russel in a lecture, 1929)
			\end{flushright}
			
			\bibliographystyle{amsalpha}

			\end{document}